\documentclass[aip,reprint]{revtex4-1}

\draft 

\usepackage{graphicx}
\usepackage{color,xspace,multirow}
\usepackage{amsmath,amssymb}
\usepackage{ulem}
\usepackage{textcomp}
\usepackage{verbatim}
\usepackage{hyperref}

\begin{document}

\title{Integrated Fiber-Mirror Ion Trap for Strong Ion-Cavity Coupling} 

\author{B.~Brandst\"{a}tter}
\email[Author to whom correspondence should be addressed. Electronic mail: ]{birgit.brandstaetter@uibk.ac.at}
\affiliation{Institut f\"{u}r Experimentalphysik, Universit\"{a}t Innsbruck, 6020 Innsbruck, Austria}
\author{A.~McClung}
\affiliation{Institut f\"{u}r Experimentalphysik, Universit\"{a}t Innsbruck, 6020 Innsbruck, Austria}
\affiliation{Norman Bridge Laboratory of Physics 12-33, California Institute of Technology, Pasadena, California 91125, USA}
\author{K.~Sch\"{u}ppert}
\affiliation{Institut f\"{u}r Experimentalphysik, Universit\"{a}t Innsbruck, 6020 Innsbruck, Austria}
\author{B.~Casabone}
\affiliation{Institut f\"{u}r Experimentalphysik, Universit\"{a}t Innsbruck, 6020 Innsbruck, Austria}
\author{K.~Friebe}
\affiliation{Institut f\"{u}r Experimentalphysik, Universit\"{a}t Innsbruck, 6020 Innsbruck, Austria}
\author{A.~Stute}
\affiliation{Institut f\"{u}r Experimentalphysik, Universit\"{a}t Innsbruck, 6020 Innsbruck, Austria}
\author{P.O.~Schmidt}
\affiliation{QUEST Institute for Experimental Quantum Metrology, Physikalisch-Technische Bundesanstalt, 38116 Braunschweig, Germany}
\affiliation{Institut f\"{u}r Quantenoptik, Leibniz Universit\"{a}t Hannover, 30167 Hannover, Germany}
\author{C.~Deutsch}
\affiliation{Laboratoire Kastler Brossel, ENS/UPMC-Paris 6/CNRS, 24 rue Lhomond, F-75005 Paris, France}
\affiliation{Menlo Systems GmbH, Am Klopferspitz 19a, 82152 Martinsried, Germany}
\author{J.~Reichel}
\affiliation{Laboratoire Kastler Brossel, ENS/UPMC-Paris 6/CNRS, 24 rue Lhomond, F-75005 Paris, France}
\author{R.~Blatt}
\affiliation{Institut f\"{u}r Experimentalphysik, Universit\"{a}t Innsbruck, 6020 Innsbruck, Austria}
\affiliation{Institut f\"{u}r Quantenoptik und Quanteninformation, \"{O}sterreichische Akademie der Wissenschaften, Otto-Hittmair-Platz 1, A-6020 Innsbruck, Austria. }
\author{T.E.~Northup}
\affiliation{Institut f\"{u}r Experimentalphysik, Universit\"{a}t Innsbruck, 6020 Innsbruck, Austria}

\date{\today}

\begin{abstract}
We present and characterize fiber mirrors and a miniaturized ion-trap design developed to integrate a fiber-based Fabry-Perot cavity (FFPC) with a linear Paul trap for use in cavity-QED experiments with trapped ions. Our fiber-mirror fabrication process not only enables the construction of FFPCs with small mode volumes, but also allows us to minimize the influence of the dielectric fiber mirrors on the trapped-ion pseudopotential. We discuss the effect of clipping losses for long FFPCs and the effect of angular and lateral displacements on the coupling efficiencies between cavity and fiber. Optical profilometry allows us to determine the radii of curvature and ellipticities of the fiber mirrors.
From finesse measurements we infer a single-atom cooperativity of up to $12$ for FFPCs longer than $200~\mu$m in length; comparison to cavities constructed with reference substrate mirrors produced in the same coating run indicates that our FFPCs have similar scattering losses. We characterize the birefringence of our fiber mirrors, finding that careful fiber-mirror selection enables us to construct FFPCs with degenerate polarization modes. As FFPCs are novel devices, we describe procedures developed for handling, aligning and cleaning them. We discuss experiments to anneal fiber mirrors and explore the influence of the atmosphere under which annealing occurs on coating losses, finding that annealing under vacuum increases the losses for our reference substrate mirrors. X-ray photoelectron spectroscopy measurements indicate that these losses may be attributable to oxygen depletion in the mirror coating. Special design considerations enable us to introduce an FFPC into a trapped ion setup. Our unique linear Paul trap design provides clearance for such a cavity and is miniaturized to shield trapped ions from the dielectric fiber mirrors.
We numerically calculate the trap potential in the absence of fibers. In the experiment additional electrodes can be used to compensate distortions of the potential due to the fibers. Home-built fiber feedthroughs connect the FFPC to external optics, and an integrated nanopositioning system affords the possibility of retracting or realigning the cavity without breaking vacuum.
\end{abstract}


\maketitle 

\section{Introduction}

Optical cavities can enhance the interaction between matter and light. In quantum information experiments, high-finesse cavities act as an interface between stationary and flying qubits, where flying qubits connect computational nodes comprised of stationary qubits. Experimentally, atoms and photons have proven to be promising candidates for the physical implementation of stationary and flying qubits \cite{Zoller05}, respectively. The building blocks for an elementary quantum network have recently been demonstrated using single neutral atoms in high-finesse cavities \cite{Ritter12}.

Trapped ions are promising candidates for quantum information processing, as techniques for high-fidelity quantum operations are well established \cite{Leibfried03a, Haeffner08}. Furthermore, trapped ions offer a range of advantages for quantum networks: ions are stably trapped in Paul traps for up to several days and can be well localized via ground-state cooling. This localization allows the ions to be accurately positioned inside a cavity mode for optimized coupling  \cite{Stute12a}.  Several research groups are currently working on the technological challenge of integrating ions and cavities, and the coupling of ions to the field of a high-finesse cavity has already been shown in a range of setups \cite{Guthoehrlein01, Mundt02, Russo09, Leibrandt09, Herskind09, Sterk12}. Single photons have been produced with a trapped ion inside a cavity \cite{Keller04, Barros09}, and entanglement between single ions and single cavity photons has recently been demonstrated \cite{Stute12}.

For a high-fidelity ion-photon quantum interface, the coherent coupling strength must be larger than the ion's rate of spontaneous decay. In order to maximize this coupling in cavity-QED systems, both the cavity length and the cavity waist should be minimized. Such microcavities require mirror surfaces with small radii of curvature. Additionally, the mirror substrates should have low surface roughness to minimize losses in the mirror coating. These low-loss coatings at optical wavelengths are dielectric. For an ion in vacuum close to such dielectric mirrors, image charges and charge build-up on the mirrors potentially distort the ion's trapping potential.

The term fiber-based Fabry-Perot cavity (FFPC) describes two opposing optical fiber tips, each with a mirror coating. At least one of the mirror surfaces is concave and aligned relative to the other such that a stable standing wave forms between them. FFPCs provide both a high coupling rate and a small dielectric cross-section and are thus a promising way to integrate ions with cavities \cite{Hunger10}.
A recent experiment provides the first demonstration of an ion coupled to a fiber-cavity mode \cite{Steiner13}.

In Sec.~\ref{sec:fiber_based_fabry_perot_resonators}, we discuss the development and optimization of FFPCs for ion traps. We develop solutions for specific technological challenges in Sec.~\ref{Sec:Annealing_Mirrors}, including the annealing and baking of fiber mirrors. Finally, in Sec.~\ref{Sec:experimental_apparatus}, we present a novel design for a linear Paul trap integrated with an FFPC. This experimental system should enable us to reach the strong coupling regime \cite{Kimble08a} with a single calcium ion inside the high-finesse FFPC.

\section{Fiber-based Fabry-Perot cavities}
\label{sec:fiber_based_fabry_perot_resonators}

Microfabricated optical cavities have several advantages over conventional cavities in cavity-QED experiments. Microcavities offer access to smaller mode volumes than have been demonstrated with macroscopic mirrors \cite{Hunger10} and thus higher interaction rates of atoms or solid-state emitters with photons. Additionally, photons that exit the cavity are directly coupled into a fiber. Furthermore, FFPCs provide flexibility in experimental setups, where small mirrors are often easier to implement than centimeter-scale mirrors fabricated on superpolished substrates. In the future it may be possible to integrate microcavity arrays in scalable systems for quantum information processing.

This range of advantages has motivated parallel development of microcavities using various technologies. We identify three criteria for microcavity development: (i) surfaces with small radii of curvature, (ii) surface roughness low enough so that it does not contribute appreciably to the mirror losses, and (iii) surfaces to which a low-loss mirror coating can be applied, i.e., by ion-beam sputtering. Such surfaces are produced by silicon wet-etching \cite{Trupke05}, enclosing nitrogen bubbles in borosilicate and polishing away the bubbles' upper half \cite{Cui06}, or by transferring a coating produced on a microlens onto an optical fiber \cite{Steinmetz06, Muller09}. All these approaches have been used to produce cavities with moderate finesses of up to $6\times10^3$. Recent developments in the fabrication of glass microcavities by shaping surfaces with controlled re-flow of borosilicate glass\cite{Roy11} yielded finesses of up to $3.2\times10^4$. However, the best microcavity finesse of $1.5\times10^5$ has been measured recently\cite{Muller10} with cavities constructed from coated, concave optical-fiber facets shaped by CO$_2$-laser ablation \cite{Hunger10}.

In this process, a short pulse of focused CO$_2$-laser light is absorbed in the cleaved tip of a fiber and creates a depression by locally evaporating the material. The created surface has a roughness of only $(0.2\pm0.1)$~nm \cite{Hunger10}. The parameters of the generated surface structures, such as radius of curvature and diameter of the depression, are set by the pulse duration, power, and beam waist of the CO$_2$ laser. A highly reflective coating is then applied to the shaped fiber surfaces by ion-beam sputtering in a high-vacuum environment. FFPCs produced in this way are being used in atom-chip setups, in which strong coupling to a BEC has been demonstrated \cite{Colombe07}. Currently, implementations of FFPCs with solid-state emitters \cite{Muller10}, ion traps \cite{Steiner13, Wilson11, VanDevender10}, and neutral atoms are being developed in several groups worldwide.

In this section, we present the recent development of FFPCs suitable for integration with ion traps. We show that we can produce cavities of length up to $350~\mu$m and finesse up to $1.1\times10^5$. Furthermore, we characterize the cavity losses due to surface roughness and the cavity birefringence, and we describe technologies for cavity alignment and fiber-mirror cleaning.

\subsection{Development of FFPCs for ion traps}
\label{Development_of_FFPCS_for_Ion_Traps}

For neutral atoms, short cavities are favorable as they provide a small mode volume and do not influence the trapping potential seen by the atom. To implement FFPCs with ion traps, however, a sufficient separation between the fibers and the ion is necessary so that the trapping potential is not distorted by charges on the dielectric mirrors.

We report on the construction of FFPCs suitable for ion traps and on the effects of increasing the cavity length, such as decreased finesse and coupling efficiencies between fiber mode and cavity mode.
Furthermore, we present measurements of general interest when working with FFPCs: a direct measurement of the scattering losses due to surface roughness, and a characterization of the birefringence of fiber mirrors.

\subsubsection{Construction of long FFPCs}
\label{Sec:Construction_of_Long_FFPCs}

\begin{figure}
  \begin{center}
    \includegraphics[width=8.5cm]{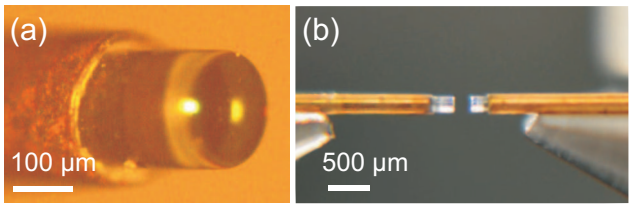}
    \caption{\label{fig:FiberCavity}(a) Composite microscope photo of a fiber mirror, assembled from multiple photos with different focal length. A highly reflective mirror coating is fabricated on the entire fiber facet. At the mirror center, a light reflection from the curved surface can be seen. (b) Photo of an FFPC. The fibers are copper coated for ultra-high--vacuum compatibility and have a cladding diameter of $200~\mu$m. The copper is etched back about $400~\mu$m from the fiber facets. The glass cladding and the gray titanium layer around the cladding, which starts about $100~\mu$m behind the facet, can be seen. The two opposing mirrors form a Fabry-Perot cavity $200~\mu$m in length.}
  \end{center}
\end{figure}

Initial development of FFPCs focused on short cavities, such as the $38.6~\mu$m FFPC used to strongly couple a BEC to a cavity field \cite{Colombe07}.
In order to construct cavities suited for ion-trap experiments, we have developed technologies that allow us to increase the length of FFPCs up to $350~\mu$m. These technologies include fabrication of structures using higher CO$_2$-laser powers and wider beam waists as well as the use of non-standard $200~\mu$m-diameter fibers.
Fig.~\ref{fig:FiberCavity} shows a composite microscope picture of a coated fiber tip and a photo of an FFPC in our laboratory.

Because of their effects on the trapping potential, fibers should remain far from the ion. As we increase the separation $L$ between the two fiber mirrors of a cavity,  the spot size of the cavity mode at each mirror increases as a function of $L$ and the mirror's radius of curvature $r$. If the mirror diameter  $2\rho_{\mathrm{m}}$ is not much larger than the field diameter $2w_{\mathrm{m}}$, the cavity mode is clipped at the mirror edge, reducing the cavity finesse.  For a conventional mirror, which has a spherical curvature over its entire surface, $2\rho_{\mathrm{m}}$ is the physical diameter of the mirror.  However, a fiber mirror can be approximated as spherical only over the length scale of the depression created in the CO$_2$-ablation process.  Thus, $2\rho_{\mathrm{m}}$ corresponds to this length scale, which is bounded above by the fiber diameter but is often much smaller due to limitations in the ablation process.

In Ref.~\cite{Hunger10}, fiber mirrors were produced with radius of curvature between $40~\mu$m and $2$~mm and $2\rho_{\mathrm{m}}$ between $10~\mu$m and $45~\mu$m.
To calculate the clipping losses associated with a particular cavity geometry, the numerical methods of Fox and Li can be used\cite{Fox61,Siegman86}.
For example, if we require round-trip clipping losses to be less than 10 ppm, then with spherical mirrors of $2$~mm radius of curvature and $2\rho_{\mathrm{m}} = 45~\mu$m, we are limited to $L \leq 70~\mu$m.  (Choosing radii of curvature in the near-confocal limit would improve this bound but also reduce the atom-cavity coupling; the near-planar assumption is a reasonable compromise and also robust to small variations in cavity length.)  This bound is incompatible with the target lengths $L \gtrsim 150~\mu$m planned for our experimental system. (See Sec.~\ref{Sec:Miniaturized_linear_Paul_trap}.)

One solution to minimize clipping losses is to produce larger mirror structures on the fiber tips, that is, to modify the laser ablation process. Specifically, we increase the beam waist at the fiber tip and use higher CO$_2$-laser power. Optimizing the laser ablation parameters is challenging due to two competing processes: while the incident laser light evaporates fiber material, mapping the Gaussian beam profile onto a concave depression, it also induces sufficient heat to locally melt the fiber tip, producing a convex structure due to surface tension. To avoid melting, heat needs to be conducted away efficiently by either cooling the fiber or creating a heat sink. Instead of standard $125~\mu$m-diameter fibers, we choose to use fibers of $200~\mu$m diameter, where the additional glass functions as a heat sink. However, the use of a non-standard fiber size means that fiber connectors and tools for cleaving and splicing are more difficult to obtain.

\begin{figure*}
  \begin{center}
    \includegraphics[width=17cm]{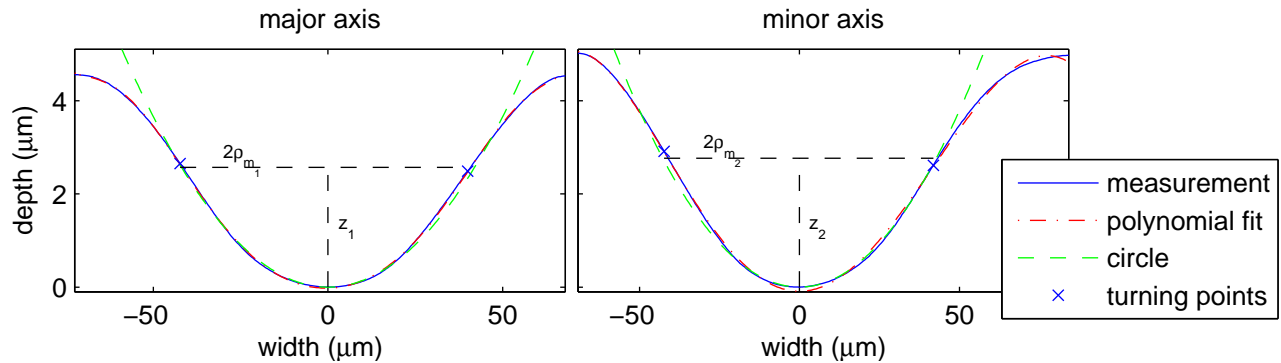}
    \caption{\label{fig:fit_fiber_surface}Fiber surface measured by optical profilometry; compare also Fig.~3 in Ref.~\cite{Hunger10}. The depression on the fiber surface is elliptical. Along the major and minor axes ($i=1,2$) of the structure, we fit a polynomial and determine structure diameter $2\rho_{\mathrm{m}_i}$ and structure depth $z_i$. $2\rho_{\mathrm{m}_i}$ and $z_i$ are defined at the turning points of the polynomial. From the fit of a circle (note the different axis scales of ordinate and abscissa), we extract the radius of curvature $r_i$. Furthermore, we determine the ellipticity of the fiber and the mean values of the fitted parameters, $r$, $2\rho_{\mathrm{m}}$, and $z$. For the surface in this figure, these values are: $2\rho_{\mathrm{m}_i}=(82,84)~\mu$m and $2\rho_{\mathrm{m}}=83~\mu$m, $z_i=(2.6, 2.8)~\mu$m and $z=2.7~\mu$m, and $r_i=(343,332)~\mu$m and $r=338~\mu$m.}
  \end{center}
\end{figure*}

Structures on the fiber tips are analyzed with an optical profilometer\cite{Fogale}, and structure diameters $2\rho_{\mathrm{m}}$ and mirror depth $z$ are extracted by fitting a polynomial to the profilometer data and finding its turning points. The distance between the turning points is defined as $2\rho_{\mathrm{m}}$. The radius of curvature $r$ of each fiber is approximated via the fit of a circle to the surface. In Fig.~\ref{fig:fit_fiber_surface}, we show the profilometer data as well as the fit for one fiber. The CO$_2$-laser ablation structures are not rotationally symmetric but have an elliptical shape due to astigmatism of the CO$_2$-laser beam \cite{Hunger10}. We determine the degree of ellipticity by identifying major and minor axes and comparing the two radii of curvature. Note that $r$, $2\rho_{\mathrm{m}}$, and $z$ are mean values of the fits to both axes.

Using the technique described here, CO$_2$-laser waists between $18~\mu$m and $80~\mu$m and powers between $0.3$~W and $1.1$~W were used in Ref.~\cite{Hunger10}. In contrast, we modify the parameters to a beam waist of $92~\mu$m and a laser power of $4.6$~W. (The pulse duration is $\sim30$~ms.) As a result, we produce fiber-mirror structures with radius of curvature between $180~\mu$m and $420~\mu$m and structure diameters of up to $80~\mu$m.

\begin{figure}
  \begin{center}
    \includegraphics[width=8.5cm]{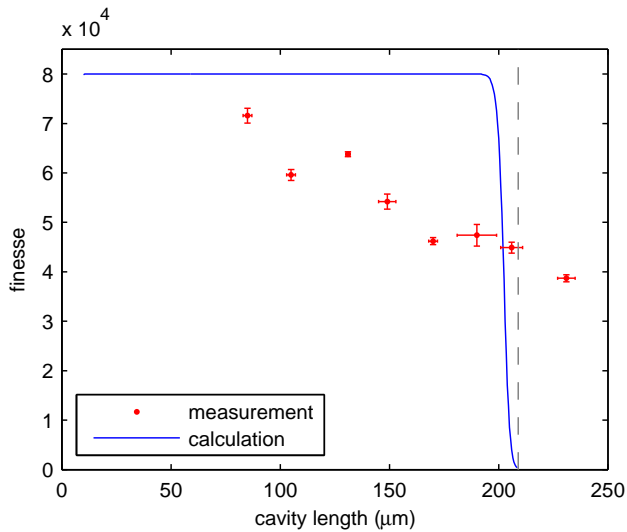}
    \caption{\label{fig:finesse_vs_length}Fiber-cavity finesse at a wavelength of $844$~nm as a function of the cavity length; compare also Fig.~$9$ in Ref.~\cite{Hunger10}. The points are measurement values from the fiber mirrors specified in the text; the finesse decreases for longer cavities. Error bars represent one standard deviation.  The solid line shows the calculated finesse due to clipping losses from the mirrors, where the mirrors are modeled as spheres with diameter 2$\rho_{\mathrm{m}}$ and radii of curvature given in the text. The grey dashed line gives the cavity's stability edge for these radii of curvature. The lack of agreement between data and calculation demonstrates that additional loss sources play a significant role.
    }
  \end{center}
\end{figure}

In a single coating process, $76$ fibers produced with the CO$_2$-laser parameters specified above were coated with a highly reflective coating centered around $860$~nm (ATFilms). On an alignment stage for test setups, we construct and characterize the FFPCs.  The mirrors were characterized at $844$~nm because a laser was available whose wavelength could be tuned over several nanometers, facilitating the measurement of short cavity lengths.  In Fig.~\ref{fig:finesse_vs_length}, we show the cavity finesse as a function of distance between the mirrors for one FFPC. 
The cavity is set up with a single-mode fiber mirror as cavity input and a multimode-fiber mirror as cavity output. The single-mode fiber mirror has a diameter of $2\rho_{\mathrm{m}}=67~\mu$m and a radius of curvature of $r=209~\mu$m; for the multimode fiber mirror, $2\rho_{\mathrm{m}}=80~\mu$m and $r=355~\mu$m.

The measured finesse declines gradually with mirror separation, from an initial value of 71,600 at $85~\mu$m to roughly half of that at $231~\mu$m.  If the decrease in finesse were due to clipping losses, we would expect a constant finesse for almost all cavity lengths, with a steep drop a few microns before the stability boundary at $r=209~\mu$m.  The clipping losses from numerical calculations, which assume spherical mirrors, are also plotted in Fig.~\ref{fig:finesse_vs_length} and do not agree with the data.

First, the most likely source of additional losses is a non-uniform thickness of the coating layers: for steep mirror surfaces, ion-beam-sputtered layers may be too thin \cite{Roy11}, shifting the coating towards wavelengths shorter than the target value. As the mode-field diameter at each mirror increases with increasing $L$, the mode may enter a region where the coating is no longer suited for the measurement wavelength and transmission losses are higher.  This effect was observed in Ref.~\cite{Roy11}, in which lower finesses for cavities with smaller radii of curvature were measured.

Second, for cavities at or beyond the stability boundary, that is, the final two data points in Fig.~\ref{fig:finesse_vs_length}, it is surprising that a nonzero finesse is observed.  For the radii of curvature determined from optical profilometry, this is not a stable cavity configuration. A possible explanation is that the assumption of a spherical mirror, while useful, is only an approximation.
Since the mode size at the mirror increases near the stability boundary, the mode extends to regions where the curvature of the mirror 
deviates significantly from a spherical fit.   A more realistic model would assume a Gaussian curvature for the mirrors, but it is difficult to use such a model to calculate the expected losses in this regime accurately; the numerical integration does not reduce to a fast Hankel transform as it does for spherical mirrors\cite{Siegman86}.

We note that Fig.~$9$ of Ref.~\cite{Hunger10} presents finesse measurements for short cavities that are compatible with the clipping-loss model.  However, for each cavity, only one data point with reduced finesse is measured, and the theory curve corresponds to an average set of parameters rather than the specific parameters of each set of mirrors.
Thus, there is not enough information to determine whether our measurements are consistent with those of Ref.~\cite{Hunger10}.  It would be interesting to explore the hypothesis of coating layer uniformity by repeating these measurements over a range of mirror curvatures.

\subsubsection{Coupling efficiencies between cavity and fiber}

In this section, we show a measurement of the transmitted intensity through an FFPC as a function of the cavity length.  The transmission is the product of four terms: the fiber in-coupling efficiency, mode matching from the fiber into the cavity, impedance matching of the cavity, and collection efficiency of the output fiber.  For free-space cavities, only the second and third terms are relevant, and the input mode is matched to the mode of the cavity by beam shaping and alignment.  With FFPCs, in contrast, the cavity mirrors are built into the in-coupling and out-coupling fibers, fixing this mode-matching coefficient to a value determined by the cavity and the fiber parameters. Because of this key difference, it is interesting to consider mode-matching from a single-mode fiber to an FFPC. 

The mode overlap $\epsilon$ is defined as the overlap between the TEM$_{00}$ mode of the cavity and the spatial mode of the single-mode fiber. This overlap depends on the radius of curvature of the mirror, on the core diameter of the fiber, and on the cavity length. In addition, either an offset of the mirror center from the fiber core or an angle between mirror and fiber core causes a mode mismatch which cannot be corrected. Ref.~\cite{Hunger10} shows that mode matching can be as high as 85\% for a short FFPC but calculates that it declines for longer cavities. Assuming that the mirror surface is orthogonal to the fiber core ($\theta=0$) and that there is no offset between the core and the mirror center ($d=0$), the coupling efficiency $\epsilon_a$ is given by \cite{Joyce84}
\begin{equation}
\label{eq:coupling_efficiency}
\epsilon_a = \frac{4}{(\frac{w_{\mathrm{f}}}{w_{\mathrm{c}}}+\frac{w_{\mathrm{c}}}{w_{\mathrm{f}}})^2+\frac{s^2}{z_\mathrm{R_f}z_\mathrm{R_c}}}
\end{equation}
with waists $w_{\mathrm{f}}$, $w_{\mathrm{c}}$ and Rayleigh lengths $z_\mathrm{R_f}$, $z_\mathrm{R_c}$ of the beam exiting the fiber and of the cavity mode, respectively. The distance from the waist of the mode exiting the fiber to the cavity waist is denoted by $s$.

A tilt of the fiber with respect to the mirror by an angle $\theta$ reduces the overlap by a factor of $e^{-(\theta/\theta_e)^2}$ for small values of $\theta$, with the angular tolerance $\theta_e$. Similarly, a displacement of the fiber core from the cavity axis reduces $\epsilon$ by a factor of $e^{-(d/d_e)^2}$, where analytic expressions for angular tolerance $\theta_e$ and displacement tolerance $d_e$ can be found in Ref. \cite{Joyce84}. The total mode overlap is then given by $\epsilon = \epsilon_a e^{-(d/d_e)^2} e^{-(\theta/\theta_e)^2} $.

\begin{figure*}
  \begin{center}
    \includegraphics[width=17cm]{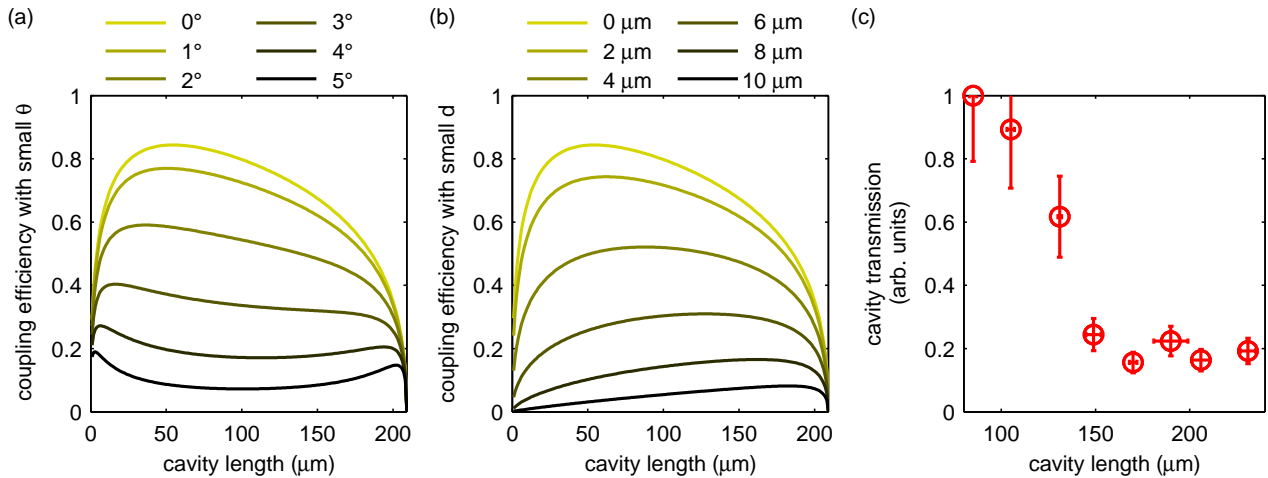}
    \caption{\label{fig:coupling_efficiency_vs_length} (a) Coupling efficiency from a single-mode fiber to an FFPC calculated from Eq.~\ref{eq:coupling_efficiency}, taking into account non-zero values of $\theta$ without displacement ($d=0~\mu$m). (b) Coupling efficiency for small displacements $d$ ($\theta=0^{\circ}$). (c) Measurement of the transmission through an FFPC as a function of the cavity length, referenced to the first data point. The error bars represent one standard deviation.}
  \end{center}
\end{figure*}

Fig.~\ref{fig:coupling_efficiency_vs_length}(a) and Fig.~\ref{fig:coupling_efficiency_vs_length}(b) show the calculated mode overlap for a single-mode fiber of $6~\mu$m core diameter and an FFPC with $r_1=209~\mu$m and $r_2=355~\mu$m.  Cavity lengths up to $209~\mu$m are plotted, considering non-zero values for $\theta$ and $d$.  For $\theta=0^{\circ}$ and $d=0~\mu$m, the mode matching between fiber and cavity increases steeply for short cavity lengths, has a maximum of $0.84$ at length $54~\mu$m, and decreases to half that value by $200~\mu$m.  Thus, we see that with proper alignment, it is possible to build long cavities with reasonable mode matching.
As $\theta$ and $d$ increase, the maximum value for $\epsilon$ drops, but $\epsilon$ becomes relatively insensitive to cavity length.  The range of values for $\theta$ and $d$ plotted in Fig.~\ref{fig:coupling_efficiency_vs_length} reflects estimates of realistic errors in the fiber-mirror fabrication procedure. Over this range, and for all cavity lengths shown, these errors cause $\epsilon$ to decrease by almost an order of magnitude.

The theory predicts a steep decrease in mode matching as the cavity length approaches the smaller radius of curvature of the two mirrors, $209~\mu$m. Therefore, when mode matching is important, cavity lengths close to the stability boundary should be avoided.  As we have seen in Sec.~\ref{Sec:Construction_of_Long_FFPCs}, however, the stability boundary of a fiber cavity does not correspond to a calculation based on spherical mirror parameters.

In Fig.~\ref{fig:coupling_efficiency_vs_length}(c), we show a measurement of the transmission of the FFPC discussed in Sec.~\ref{Sec:Construction_of_Long_FFPCs}. Note that due to the large core diameter and acceptance angle of multimode fibers, collection efficiency is unity for our cavity parameters \cite{Hunger10}.
We measure the transmission through the cavity and the fibers normalized to the transmission of the first point as a function of the cavity length.
The transmission first decreases as a function of cavity length and then remains constant at around $20\%$ of the initial value for cavities longer than $150~\mu$m.
Only the relative transmission is measured because an absolute transmission is difficult to calibrate and does not provide additional information on how mode overlap scales with length \cite{Hunger10, Hood01}.
In order to extract the mode-matching efficiency from this transmission measurement and compare it to Figs.~\ref{fig:coupling_efficiency_vs_length}(a) and (b), we would have to determine independently
whether the decrease in transmission is due to increasing impedance mismatch or increasing mode mismatch.

For experiments in which optimal transmission through long fiber cavities is important, the coupling can be improved by minimizing cavity scatter and absorption losses, therefore optimizing the impedance matching.  The mode matching can be maximized by minimizing $\theta$ and $d$.  An interesting possibility for improving the mode matching would be to tailor the single-mode fiber mode, e.g., by expanding the fiber core at the tip.

\subsubsection{Surface-loss measurement}
\label{sec:surface_loss_measurement}

We measure the scattering losses in the mirror coatings due to surface roughness of the CO$_2$-laser shaped fibers, and we find no such losses within a $1$~ppm measurement error. This measurement is the first direct comparison of the losses of fiber-mirror coatings (including losses induced by surface roughness) with losses of identical coatings fabricated on fused silica substrates. For this purpose, fibers and substrates were coated together in the same fabrication run.

The losses of highly reflective mirror coatings depend critically on the surface quality of the mirror substrate: surface roughness of the substrate material results in scattering losses of the mirror. In order to reduce the scattering losses to $\sim1$~ppm at near-infrared wavelengths, the surface roughness needs to be less than $1$~\r{A}~rms \cite{Hunger10}. Mirror substrates of surface roughness less than $1$~\r{A}~rms are referred to as superpolished. With superpolished substrates, we assume that all scattering and absorption losses come from point defects in the coating.

As CO$_2$-laser ablation is a novel technique for producing curved mirror substrates, it is of great interest to determine the quality of the shaped surfaces and the induced scattering losses. Fiber surface roughness $\sigma$ has previously been measured with an atomic force microscope and linked to the scattering losses $S$ via $S\approx(4\pi\sigma/\lambda)^2$, where $\sigma = (0.2\pm0.1)$~nm, corresponding to $S=(10\pm10)$~ppm for near-infrared light at $\lambda=780$~nm \cite{Hunger10}.

In contrast, we compare fiber mirrors with reference mirrors produced on superpolished substrates.
We measure reference-mirror cavities to have a finesse of $(1.10\pm0.04)\times10^5$, in comparison to a finesse of $(1.14\pm0.05)\times10^5$ for the fiber-mirror cavities. For identical mirrors, the total losses per mirror $\mathcal{L}_{\mathrm{tot}}=\mathcal{T+L}$ , the sum of transmission $\mathcal{T}$ and losses $\mathcal{L}$, are calculated from the finesse via $\mathcal{L}_{\mathrm{tot}}=\pi/\mathcal{F}$. The reference substrates and the fiber mirrors thus have identical total losses of $\mathcal{L}_{\mathrm{tot}}=(28\pm1)$~ppm. In an additional measurement (Sec.~\ref{sec:repeated_annealing_under_vacuum_and_under_air}), we determine the total losses of the same reference substrates after annealing the mirrors. This measurement of losses of $(17\pm1)$~ppm agrees with $15$~ppm target transmission of the coating and $2$~ppm scatter and absorption losses from the coating.

We conclude that fiber surface roughness does not cause any additional scattering losses. This result suggests that it may be possible to construct FFPCs with finesses as high as those achieved with state-of-the-art superpolished mirrors \cite{Rempe92}.

\subsubsection{Birefringence of fiber mirrors}

We investigate the birefringent splitting of orthogonal polarization modes in FFPCs. We observe that FFPCs exhibit significant birefringence, whereas cavities built with mirrors produced on superpolished substrates fabricated in the same coating process do not exhibit measurable birefringence. Furthermore, we observe that the birefringent splitting of the FFPCs varies as a function of the cavity alignment.

For experiments in which quantum information is encoded in photon polarization, it is advantageous for the modes of orthogonal linear polarization to be degenerate in the cavity~\cite{Ritter12, Stute12}. Therefore, it is of interest to control the birefringence of the cavity mirrors. Typically, the birefringence in a cavity of mirrors fabricated on superpolished substrates is induced via stress inside the mirrors due to the mounting~\cite{Lynn03a}.  In contrast, fiber mirrors are mounted a few millimeters from the mirror surface, a length scale much greater than the surface diameter. Therefore, we assume that the stress is intrinsic to the coating and is created in the fabrication process of the coating.

To measure the birefringent splitting, we measure the transmission curve of an FFPC while driving the cavity and sweeping its frequency. To calibrate the abscissa, we modulate the driving laser to produce frequency sidebands with a known splitting. We repeatedly rotate and realign the fibers and measure the relative detuning between the polarization modes. The observed splittings range from smaller than the FWHM of the cavity resonance ($ \approx  30 $~MHz) up to a few gigahertz. This result is comparable to the birefringent splitting of $200$~MHz measured in Ref.~\cite{Hunger10}. Furthermore, when aligning the cavity to support different TEM modes, we observe that these TEM modes have different birefringent splittings.

\begin{figure*}
  \begin{center}
    \includegraphics[width=17cm]{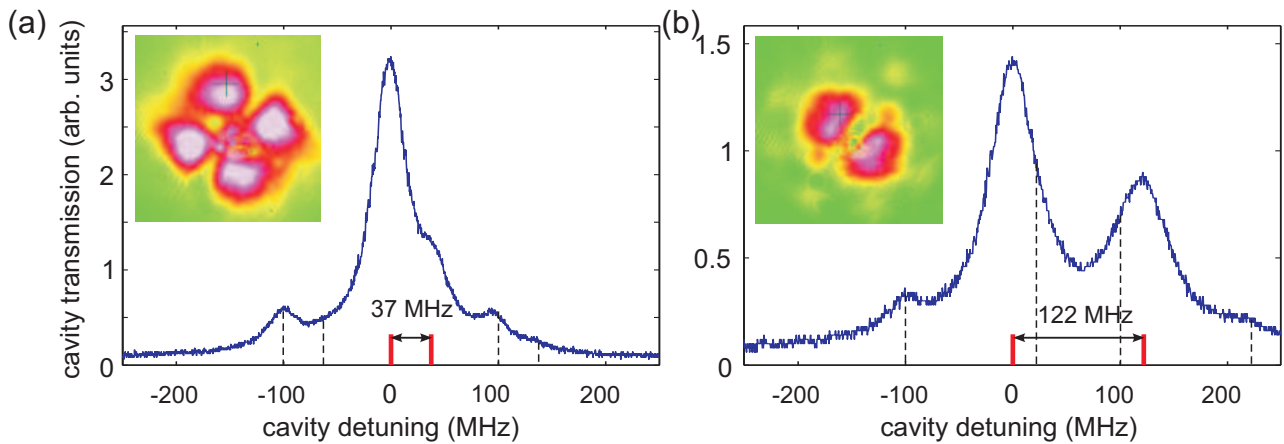}
    \caption{\label{fig:birefringence_modes} Birefringent splitting of the orthogonal polarization modes inside the cavity described in the text. Birefringent splitting of (a) $37$~MHz for TEM$_{11}$ and of (b) $122$~MHz for TEM$_{01}$. The $100$~MHz sidebands for the frequency calibration are indicated with dashed lines. The insets show CCD-camera images of the cavity modes.}
  \end{center}
\end{figure*}

To quantify this last observation, we have modified the experimental setup because the TEM mode of the cavity is not preserved inside the output multimode fiber. The multimode fiber is replaced by a mirror fabricated on a superpolished substrate that was coated in the same coating run as the fiber. The TEM mode is then imaged with a CCD camera, while a photodiode measures the cavity transmission curve. We measure a splitting of $37$~MHz for TEM$_{11}$ and $122$~MHz for TEM$_{01}$ (Fig.~\ref{fig:birefringence_modes}), demonstrating that the birefringent splitting is dependent on the TEM mode and thus on the specific mirror region that the cavity mode samples.

We conclude that birefringence is not homogeneous over the mirror coating on the fiber. Thus, simply rotating two birefringent fiber mirrors with respect to one another will not necessarily eliminate cavity birefringence. However, by careful selection of fibers and proper alignment, we have built cavities which satisfy a target birefringent splitting, in our case, degenerate polarization modes.

\subsection{Ion-cavity system}

We now focus on the relevant rates of our experimental ion-cavity system considering $^{40}\mathrm{Ca}^+$ and the FFPC parameters from the measurements shown above. Until very recently, experimental ion-cavity systems have used cavities constructed with superpolished-substrate mirrors \cite{Guthoehrlein01, Mundt02, Russo09, Leibrandt09, Herskind09}; in addition, an ion has now been coupled to an FFPC \cite{Steiner13}.
All of these systems operate in a regime in which the coherent coupling rate $g$ between a single ion and a photon is smaller than the rate of at least one incoherent process, such as scattering from the ion or loss from the cavity.
To increase the coherent coupling strength $g$ between ion and photon, the cavity mode waist $w_0$ and the cavity length $L$ should be minimized, as for a dipole coupling
\begin{equation}
g = \frac{\lambda}{\pi w_0}\sqrt{\frac{3 c \gamma_c}{L}},
\end{equation}
where $\gamma_c$ is the spontaneous emission rate between the states coupled via the cavity, and $w_0$ is calculated via the length $L$ and the radius of curvature $r$ of the mirrors. For a cavity in which both mirrors have the same radius of curvature,
\begin{equation}
w_0^2=\frac{\lambda}{2\pi} \sqrt{L(2r-L)}.
\end{equation}

Laser machining of fiber mirrors produces radii of curvature that are two orders of magnitude smaller than radii of curvature produced via superpolishing techniques \cite{Hunger10}. Thus, short cavity lengths and small cavity waists are inherent to FFPCs, allowing for high coupling rates.
Our setup is comprised of a linear Paul trap, described in Sec.~\ref{Sec:Miniaturized_linear_Paul_trap}, and an FFPC. In the setup, the cavity axis is perpendicular to the trap axis. Ions trapped along this axis would have a separation of around $5~\mu$m. Thus, the number of ions that can be coupled to the same antinode of the cavity depends on the size of the cavity waist. For a mode waist $w_0$ of $5~\mu$m, it is possible to have two ions displaced symmetrically from the maximum of the cavity mode. Both ions are then coupled to the cavity with $88\%$ of the maximum strength.

Our cavity mirrors are chosen for optimum finesse at either the $4P_{1/2}-3D_{3/2}$ or the $4P_{3/2}-3D_{5/2}$ transition of $^{40}\mathrm{Ca}^+$, which have wavelengths of $866$~nm and $854$~nm, respectively. The $3D_{3/2}$ and $3D_{5/2}$ states are metastable states, and the $3D_{5/2}$ state is used for quantum-information processing \cite{Haeffner08}.
Here, we calculate the system parameters, that is, the coherent coupling rate $g$ and cavity decay rate $\kappa$, for the FFPC characterized in Sec.~\ref{Sec:Construction_of_Long_FFPCs}. The finesse and the cavity length have been measured, and $w_0$ and $g$ are calculated from the radius of curvature measured interferometrically and are therefore approximate values. $\kappa$ is calculated via the finesse $\mathcal{F}$ by
\begin{equation}
\kappa = \frac{c \pi}{2\mathcal{F} L},
\end{equation}
and the single-atom cooperativity $C$ is given by
\begin{equation}
C = \frac{g^2}{2\kappa\gamma}.
\end{equation}

\renewcommand{\arraystretch}{1} 
\setlength{\tabcolsep}{8pt} 

\begin{table*}[bt]
\caption{\label{Tab:system_parameters} Cavity-QED system parameters for various setups of neutral-atom and ion experiments using FFPCs and cavities consisting of superpolished mirror substrates. These parameters are compared to our $^{40}\mathrm{Ca}^+$--FFPC system with the FFPC characterized in Sec.~\ref{Sec:Construction_of_Long_FFPCs}; given the measured values of cavity length $L$ and finesse $\mathcal{F}$, we calculate the mode waist $w_0$, cavity decay rate $\kappa$, coherent coupling rate $g$ and single-atom cooperativity $C$.}
\begin{tabular}{crrrcccc}
\hline \hline
experiment & $L$ ($\mu$m) & $w_0$ ($\mu$m) & \multicolumn{1}{c}{$\mathcal{F}$} & $\kappa$ \newline (MHz$/2\pi$)  & $g$ (MHz$/2\pi$) & $\gamma$ (MHz$/2\pi$) & $C$\\
\hline
\hline
neutral $\text{Cs}$ \cite{Hood00a} & $10.9$ & $14.0$ & $480,000$ & $14.1$ & $110$ & $2.6$ & $164$\\
\hline
neutral $\text{Rb}$ - FFPC \cite{Colombe07} & $38.6$ & $3.9$ & $37,000$ & $53$ & $215$ & $3$ & $145$\\ 
\hline
\hline
$\text{Ca}^+$ ions \cite{Keller04} & $8000.0$ & $37.0$ & $78,000$ & $1.2$ & $0.92$ & $11.2$ & $0.03$\\ 
\hline
$\text{Ca}^+$ ions \cite{Stute12a} & $19960.0$ & $13.2$ & $77,000$ & $0.05$ & $1.43$ & $11.2$ & $1.8$\\ 
\hline
$\text{Yb}^+$ ions - FFPC \cite{Steiner13} & $230.0$ & $6.6$ & $1,000$ & $320$ & $6$ & $2$ & $0.03$\\ 
\hline
\hline
current setup & $85.0$ & $5.1$ & $72,000$ & $12$ & $41$ & $11.2$ & $6.1$\\
$\text{Ca}^+$ ions - FFPC & $131.0$ & $5.4$ &  $64,000$ & $9$ & $31$ & $11.2$ & $4.8$\\
 & $206.0$ & $3.2$ &  $45,000$ & $8$ & $41$ & $11.2$ & $9.3$\\
\hline
\end{tabular}
\end{table*}

The parameters of our ion-cavity system are compared in Tab.~\ref{Tab:system_parameters} to a selection of single-atom cavity-QED experiments. 
With neutral Cs and Rb atoms, cooperativities over $10^{2}$ have been demonstrated by using short, high-finesse cavities \cite{Hood00a} and by using an FFPC to obtain a small mode waist \cite{Colombe07}.  In contrast, ion-trap experiments have been limited to $C \lesssim 1$, primarily because relatively long cavities have been necessary to avoid distortion of the trapping potential \cite{Keller04,Stute12}.  Here, we see that fiber mirrors offer a promising route towards much shorter cavities and smaller mode waists, as in the first demonstration of an ion-trap FFPC \cite{Steiner13}.

In our system, the atomic decay rates $\gamma$ of the $4P_{3/2}$ state and the $4P_{1/2}$ state are $2\pi\times 11.4$~MHz and $2\pi\times 11.2$~MHz, respectively, including decay channels to both $S$ and $D$ manifolds. From Tab.~\ref{Tab:system_parameters}, we see that the coherent coupling rate $g$ is larger than $\kappa$ and $\gamma$ for the range of possible cavity lengths of this FFPC. Longer cavities exhibit smaller $\kappa$, although the finesse decreases with increasing cavity length (Fig.~\ref{fig:finesse_vs_length}). Additionally, the sharp decrease in cavity waist gives increasingly larger $g$ as the near-concentric limit is approached. These trends contribute to higher cooperativities as the cavity length is increased.

\subsection{Practical techniques for FFPCs}

FFPCs differ from conventional cavities in several ways. The mirror is fabricated on the fiber, so that different techniques are required to align a cavity. Furthermore, due to the small size of both fibers and mirrors, technologies for mounting and cleaning as well as for annealing of the fiber mirrors are necessary. In this section, we describe new methods developed in our work with FFPCs. Experiments annealing fiber mirrors are presented separately in Sec.~\ref{Sec:Annealing_Mirrors} together with measurements of annealing mirrors fabricated on superpolished substrates.

\subsubsection{Fiber preparation}

We work with copper-coated single-mode and multimode fibers of non-standard $200~\mu$m diameter\cite{IVG}. Here, we summarize techniques for fiber preparation.

{\bf Etching copper coating from fibers:} Fibers need to be stripped properly before they are cleaved or connectorized. We etch away the copper with a $25\%$ nitric acid (HNO$_3$) solution or a $20\%$ iron(III) chloride (FeCl$_3$) solution at $50^{\circ}$C until the coating is no longer visible. This process takes only a few minutes. The first method is faster but also requires more precaution in handling the chemicals.

{\bf Removing titanium from fibers:} After the copper is etched off, the glass fibers are still coated with a thin layer of titanium, which is used as an adhesive between glass and copper. This layer is not insulating, and if the fibers have contact with trap electrodes, they cause a short circuit. Furthermore, the titanium layer shifts the ion's trapping potential when it is brought close to the trap center. We find that the titanium can be scratched away gently with diamond paste, which is then rinsed off thoroughly with solvents.

{\bf Mounting fibers:} To protect the fibers from dirt or damage, they need to be mounted properly during every stage of the experimental process, e.g., in the coating chamber, during testing or in the experimental setup. In the coating device, the fiber holders need to be vacuum compatible. We clamp each fiber with a screw inside an aluminum cylinder. The fiber tip protrudes $0.5$~mm from the cylinder for the coating. We encase the fiber in a Teflon sleeve so that the screw does not damage it. The same holders are used for fiber storage before and after coating. However, to set up a test cavity, the fibers should not be clamped rigidly with screws as it impairs their optical transmission. Instead, we fix the fibers with a magnet in stainless-steel v-grooves. In the experimental setup, the fibers are then glued with UHV epoxy\cite{EPOTEK} onto Pyrex v-grooves.

{\bf Connectorizing fibers:} In the testing process, we often switch between fibers and thus prefer slide-on slide-off bare-fiber adapters to connectors that need to be glued. The bare-fiber adapters are custom-made for the $200~\mu$m cladding diameter\cite{Bullet}.

{\bf Splicing non-standard $200~\mu$m diameter fibers to standard $125~\mu$m diameter fibers:} To be able to use standard fiber tools, it is useful to work with $125~\mu$m diameter fibers. We find that it is possible to splice the $200~\mu$m diameter fibers to standard $125~\mu$m diameter fibers of the same core diameter with a commercial fiber splicer\cite{Vytran} with negligible losses. In the splicing process, we account for the larger diameter of the $200~\mu$m fiber by shifting the splice filament such that it preferentially heats the larger fiber.

\subsubsection{Alignment of long cavities}

To obtain a cavity-transmission signal, it is sufficient to align the fibers by eye to form a very short cavity of about $30~\mu$m in length and sweep the length across a free spectral range. However, as the CO$_2$-laser ablation does not produce perfect surfaces --- generally, the center of the depression is offset from the center of the fiber facet, and the mirror surface is not exactly orthogonal to the fiber core --- the cavity must be aligned further via optimization of the cavity transmission signal. For aligning FFPCs, one fiber is fixed while the second fiber is mounted on a six-axis nano-positioning system\cite{Thorlabs}. To build longer FFPCs, the distance between the mirrors is then increased stepwise while optimizing the alignment by maximizing the transmission signal. Using this technique, we are able to build cavities of lengths up to $350~\mu$m.

We observe that cavities of up to $\sim 100~\mu$m in length are robust to changes in mirror position or angle. As the fiber mirrors are separated further, however, very small changes misalign the cavity even though the mirrors' radii of curvature suggest that the cavity is still far away from the edge of the stability region. In this case, the size of the mirror is the limiting factor for misalignment. As a consequence, for building long cavities, care must be taken that the fibers are mounted very stably.

\subsubsection{Cleaning fiber mirrors}

To clean ultra-low loss mirrors fabricated on a superpolished substrate, the mirror is rotated on a spin cleaner and the surface is swabbed with water, acetone and isopropyl alcohol during rotation \cite{Northup08a}. Fiber mirrors, however, are too delicate to swab. The high-temperature gradient of the CO$_2$-ablation process makes the fiber tip brittle, and we find that any stress or pressure usually breaks the tip.

One obvious strategy has been to keep the fiber mirrors as clean as possible and shield them from contamination. Unfortunately, even in a clean environment, the fiber mirrors sometimes decrease in finesse as they accumulate dust. To address this problem, we have developed a cleaning procedure for ultra-low loss fiber mirrors. We use spectrophotometric grade solvents, heated to $50^{\circ}$C, to clean the mirror fiber tips in an ultrasonic bath for two minutes, first in acetone and then in methanol. Immediately after taking the fiber out of the methanol, we use clean helium to dry the mirror surface for at least half a minute.

There are a few cases in which this method does not recover the initial finesse. However, we typically see full recovery of the finesse by cleaning fiber mirrors with this procedure.

\section{Annealing mirrors}
\label{Sec:Annealing_Mirrors}

Ultralow-loss mirrors at optical wavelengths are routinely employed in quantum optics experiments. Using ion-beam sputtering, mirrors can be fabricated with total losses (transmission, absorption, scatter) as low as $1.6$~ppm in the near infrared \cite{Rempe92}. In order to achieve such low losses in dielectric mirror coatings, it is a standard procedure to anneal the coatings after fabrication. Annealing leads to homogenization of the oxide layers and improves the stoichiometry of non-perfect oxides \cite{Atanassova95}, reducing coating losses typically by $10$~ppm. This procedure is thus a key step in the process of manufacturing ultra-low-loss mirrors.

The recent development of fiber cavities raises the question of whether annealing fiber mirrors is possible. Since the surface roughness of CO$_2$-laser--ablated fiber tips is comparable to that of superpolished mirror substrates (Sec.~\ref{sec:surface_loss_measurement}), the finesse of fiber cavities can in principle reach the record values achieved with mirrors fabricated on substrates. To reach this high finesse, however, annealing would be essential.

Our initial efforts to anneal fiber mirrors have been unsuccessful. We attribute some of the difficulties to possible chemical reactions of the fiber-coating material with oxygen in the air. For this reason, we have investigated annealing under vacuum. Furthermore, knowledge about the effects of baking mirrors under vacuum is essential for all experiments in which low-loss mirrors are placed under ultra-high vacuum (UHV), which requires a vacuum bake. The typical temperatures for a vacuum bake are lower than annealing temperatures, but the same chemical processes are at work. In various experiments, degradation of cavity mirrors under vacuum has been observed (Ref.~\cite{Cetina13, Sterk12} and references therein), but evidence of changes in mirrors under vacuum has been mostly anecdotal. Because the cavities are part of complex experimental systems, in which repeated bake-outs are impractical, a careful study of these effects has not yet been undertaken.

In order to study annealing and baking under vacuum systematically, we use reference mirrors which have been produced in the same coating run as the fiber mirrors (Sec.~\ref{Development_of_FFPCS_for_Ion_Traps}). These reference mirrors are fabricated on fused silica substrates of half-inch and $7.75$~mm diameters. They are coated with a highly reflective coating comprised of $37$ alternating layers of Ta$_2$O$_5$ and SiO$_2$, where the inner- and outermost layers are Ta$_2$O$_5$. The layers are deposited using ion-beam sputtering, and each layer has a $\lambda/4$ thickness for peak reflectivity at $\lambda = 860$~nm. Using these mirrors, we systematically measure effects from annealing under air and vacuum in a clean and controlled system.

In this section, annealing refers to a $90$~minute ramp from room temperature to $450^{\circ}$C, a $90$~minute bake at $450^{\circ}$C, and a $90$~minute ramp down to room temperature. 
Vacuum annealing consists of placing the mirrors in a clean stainless-steel chamber, which is then pumped to pressures lower than $10^{-5}$~mbar by a turbo pump, after which the temperature ramp is started. For annealing under air, the mirrors are placed inside clean glass Petri dishes. Care is taken that the mirrors are properly cleaned before any finesse measurement \cite{Northup08a}.

\subsection{Repeated annealing under vacuum and under air}
\label{sec:repeated_annealing_under_vacuum_and_under_air}

\begin{figure}
    \includegraphics[width=8.5cm]{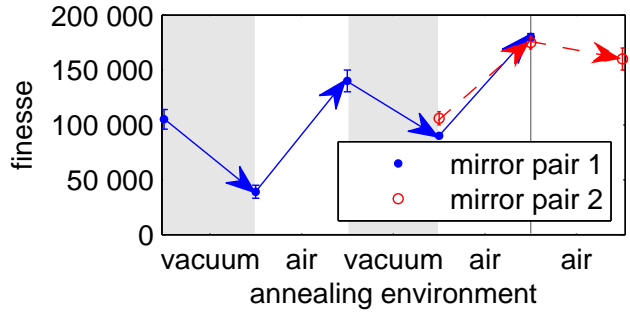}
    \caption{ \label{fig:bakeCycle}Finesse after annealing at $450^{\circ}$C under alternating air and vacuum environments. Annealing under vacuum shows repeatable losses in cavity finesse, and annealing under air repeatable gains up to a maximum finesse of $(1.80\pm0.03)\times 10^5$ for mirror pair $1$ (points). Mirror pair $2$ (open circles) establishes the maximum finesse after repeated annealing under air. The error bars represent one standard deviation of the measurement uncertainty.}
\end{figure}

With a first pair of coated mirrors, we constructed a cavity with a finesse of $(1.05\pm 0.09)\times 10^5$ prior to annealing. After an initial test in which the mirror pair was annealed under vacuum, the finesse had degraded to $(3.9\pm 0.6)\times 10^4$. To investigate this unexpected result, we conducted a series of measurements, in which we alternated between annealing under vacuum and air. Annealing these mirrors under air resulted in a recovery of the finesse, that is, a decrease of the losses that had been induced by vacuum annealing. In fact, the new finesse of $(1.4\pm 0.1)\times 10^5$ was higher than the initial value, indicating that annealing had removed intrinsic coating losses as expected. Two subsequent measurements showed that the losses when annealing under vacuum and gains when annealing under air are repeatable, and the maximum finesse for this pair of mirrors is $(1.80\pm0.03)\times 10^5$; these data are summarized in Fig. \ref{fig:bakeCycle}.

With a second pair of reference mirrors, we reproduce the initial finesse of the first pair: $(1.06\pm 0.06)\times 10^5$. Annealing directly under air as the only step yields a finesse of $(1.75\pm 0.06)\times 10^5$, implying that the maximum finesse of this coating is independent of previous annealing cycles. Repeated annealing under air established the maximum finesse for this pair of mirrors. This finesse corresponds to total losses $\mathcal{L}_{\mathrm{tot}}$ of $17$~ppm. We attribute 2ppm \cite{Rempe92} to scattering and absorption losses and $15$~ppm to transmission, consistent with the target transmission of the coating run.

Our finding that the change in finesse depends on the annealing environment suggests that annealing affects the chemical composition of the mirror coating. We hypothesize that during a vacuum bake, oxygen escapes from the outermost Ta$_2$O$_5$ layer, which leads to defects in the coating. Subsequent annealing under air gives the surface the possibility to regain the oxygen, thus removing these defects of the chemical structure. In order to test this theory of oxygen depletion, we conducted a series of X-ray photoelectron spectroscopy measurements.

\subsection{X-ray photoelectron spectroscopy measurements}

X-ray photoelectron spectroscopy (XPS) is used to quantitatively determine the chemical composition of the Ta$_2$O$_5$ layer on the surface of the mirror coatings. We acquire the XPS data with a Thermo Multilab $2000$ utilizing monochromatic Al~K$\alpha$ radiation at $1486.6$~eV. The atomic composition of the samples is obtained from XPS survey scans taken with an overall resolution of $2.0$~eV. The oxygen and tantalum content are determined from the O ($1$s) and the Ta ($4$d) lines, respectively, measured with a higher resolution of $0.1$~eV. Measured intensity ratios are converted into atomic percentages using the theoretical photoionization cross-sections of Scofield \cite{Scofield76}, also taking into account the energy-dependent transmission of the electron-energy analyzer \cite{Klauser10}. 

\subsubsection{XPS measurements of mirrors annealed under air and under vacuum}
\label{sec:Comparison_of_Mirror_Annealed_Under_Air_and_Under_Vacuum}

We acquire XPS spectra from two mirrors and compare their chemical composition. Prior to the measurement, one mirror was annealed in air, while the other was annealed in vacuum. To calculate the amount of oxygen and tantalum from the XPS spectra, we subtract the background and integrate over the O ($1$s) and Ta ($4$d$5$) photoelectron lines. When weighted by the Scofield sensitivity factors, which represent the emission probability of an electron, these integrals give the relative proportions of the elements in the material. The sensitivity factor is $15.64$ for the Ta ($4$d$5$) line and $2.93$ for the O ($1$s) line.

Using this method, we compare the chemical compositions of the two mirrors. The oxygen concentration of the mirror annealed under vacuum is $(0.9\pm0.7)\%$ lower than the oxygen concentration of the mirror annealed under air. This difference would constitute a loss of every $90$th oxygen atom from the surface of the mirror annealed under vacuum. The large error bars are due to the relative uncertainty of $0.5\%$ between measurements on the same apparatus. To resolve the effect more clearly, we conduct a second experiment based on the measurement of a single mirror over time.

\subsubsection{Continuous XPS measurement during vacuum annealing}
\label{sec:Continuous_XPS_Measurement}

\begin{figure}
  \includegraphics[width=8.5cm]{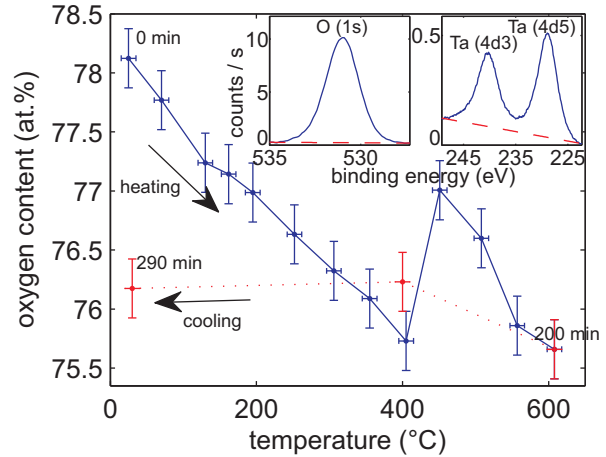}
  \caption{\label{fig:XPS_spectrum_continuous}
  Oxygen content in the Ta$_2$O$_5$ layer on the mirror surface. The temperature of the mirror coating is increased stepwise and XPS measurements are taken at each temperature. The mirror has been annealed in air before the measurement. The solid line is taken during the heating process, the dotted line during the cool-down of the substrate. The uncertainty of the temperature measurement is $20^{\circ}$C. The relative uncertainty of the oxygen content is $0.5$\%. The insets show sample XPS spectra of the O (1s) and the Ta (4d) lines (solid), including the background (dashed), at one temperature setting.}
\end{figure}

To observe the effect of oxygen loss from the surface directly, we perform real-time XPS measurements during the process of annealing in vacuum on a mirror that has previously been annealed in air. The mirror is placed inside the XPS vacuum chamber and a reference XPS measurement is taken. Between subsequent XPS measurements, the mirror temperature is increased stepwise up to $608^{\circ}$C over $200$~minutes. This procedure gives an exact chemical analysis of the mirror surface at each step of the annealing process. Integrating the area under the oxygen and tantalum peaks, we calculate the oxygen content in the surface of the mirror. The insets of Fig.~\ref{fig:XPS_spectrum_continuous} show the O ($1$s) and the Ta ($4$d) lines of one of the XPS spectra which we use for this analysis. The temperature of the mirror in this measurement is measured with a pyrometer (IMPAC) with a measurement uncertainty of $20^{\circ}$C.

Figure~\ref{fig:XPS_spectrum_continuous} shows the results of these measurements, in which the oxygen content decreases as the temperature increases. The atomic percentage of oxygen of the mirror before annealing is $78.2$\%; at $405^{\circ}$C it drops to $75.7$\%. The oxygen content briefly recovers between $450^{\circ}$C and $550^{\circ}$C, suggesting a phase transition \cite{Bansal94} or outgassing of oxygen. At $608^{\circ}$C, the oxygen content reaches its lowest point of $75.7$\%. The mirror cool-down lasts $90$ minutes, during which the oxygen content neither decreases nor increases significantly. We note that the absolute uncertainty of the measurement is around $10$\%. The entire annealing process results in a $(2.5\pm0.7)\%$ drop in oxygen content, supporting the hypothesis of oxygen depletion from the Ta$_2$O$_5$ layer.

The discrepancy between this result and our earlier measurement of Sec.~\ref{sec:Comparison_of_Mirror_Annealed_Under_Air_and_Under_Vacuum} might be due to oxygen reuptake when the vacuum-baked mirror was in air before the XPS measurement. Both measurements show that we can attribute the lower finesse of the vacuum-annealed mirrors to the loss of oxygen of the Ta$_2$O$_5$-layer on the mirror surface.

\subsection{Baking under vacuum}

Up to now, we have only presented measurements of mirror annealing at temperatures of $450^{\circ}$C and higher. The depletion of oxygen observed at these temperatures suggests that this effect also takes place --- in a moderate form ---  when baking mirrors at standard temperatures for a vacuum bake-out, typically $200^{\circ}$C to $300^{\circ}$C. The XPS measurement of Sec.~\ref{sec:Continuous_XPS_Measurement} shows a linear decrease of oxygen when heating the mirror from room temperature up to $405^{\circ}$C in vacuum. At a temperature of $160^{\circ}$C one percent of the oxygen is already lost from the surface, and at $300^{\circ}$C $1.8\%$ of the oxygen is lost.

We expect that if we bake mirrors under vacuum conditions at different temperatures, one would find decreasing mirror finesses as the temperature of the bake increases. This measurement would presumably show the same temperature dependence of oxygen depletion following a vacuum bake as the XPS measurements. We can estimate the mirror losses by linearly extrapolating the two annealing measurements from Fig.~\ref{fig:bakeCycle} to lower temperatures. The first annealing under vacuum was performed with non-annealed mirrors, while the second time, these mirrors had been pre-annealed under air. According to these measurements, we would expect $32$~ppm and $8$~ppm of additional losses at a baking temperature of $300^{\circ}$C for the non-annealed mirrors and the annealed mirrors, respectively.
To understand whether this difference in losses can be attributed to the pre-annealing, it would be interesting to repeat the measurements for several mirror pairs.

A pre-annealed test mirror, however, baked under vacuum conditions at $300^{\circ}$C, showed a decrease of the finesse from $1.9\times10^5$ to $4\times10^4$ after baking. These $62$~ppm of additional mirror losses are higher than expected from the annealing results. Furthermore, the mirror finesse could not be recovered by successive air bakes, suggesting that the mirror was damaged in the baking process.  Lacking additional undamaged test mirrors, we did not perform further vacuum bakes at moderate temperatures.  However, a systematic study of vacuum bakes over a range of temperatures would provide valuable information for cavity setups in UHV.

\subsection{Discussion}

Annealing under vacuum decreases the mirror finesse rather than increasing it. As a consequence, fiber mirrors should not be annealed under vacuum.

Tests of annealing fiber mirrors under air have not been successful so far. Even when a clean annealing environment is established, the fiber mirrors seem to get dirty after baking in air and cannot be cleaned successfully. We suspect that contamination from the copper coating of the fiber damages the mirror coating. A way to remove this source of contamination is the use of chemically more inert fiber coating materials such as gold.

Furthermore, we expect the mirror losses to increase under a vacuum bake even at moderate temperatures. The XPS measurements of Sec.~\ref{sec:Continuous_XPS_Measurement} show that the oxygen decreases linearly with increasing baking temperature, and thus the amount of defects in the mirror increases. Vacuum baking should therefore be done at the lowest possible temperatures, although baking at higher temperatures under oxygen atmosphere might be a solution.

\section{Experimental ion-trap apparatus with an FFPC}
\label{Sec:experimental_apparatus}

In our combined FFPC ion-trap setup, the fibers sit on UHV-compatible positioners enabling in-vacuum cavity alignment. In addition, these positioners allow us to pull back the fibers from the trap center, so that the ion trap can be tested without the influence of the fibers on the ion's trapping potential. The ion trap is a modified version of the linear Paul trap presented in Refs.~\cite{Gulde03a, Riebe05a}. Here, we describe in detail the ion trap, the integration of the FFPC into the trap setup, and the underlying design considerations.

\subsection{Experimental design considerations}

Ions are trapped quasi-permanently and well isolated from environmental perturbations in RF Paul traps \cite{Paul90} under UHV conditions. 
Possible ion-trap designs include surface-electrode traps \cite{Chiaverini05}, segmented linear traps based on microchip technology \cite{Schulz08}, `endcap' \cite{Schrama93, Wilson11} or stylus ion traps \cite{Maiwald09}, and linear blade traps \cite{Gulde03a, Riebe05a}. Two criteria for selecting a specific design are low ion heating rate and deep trap depth, both of which contribute to long ion lifetimes in the trap. The heating rate increases with decreasing ion-electrode distance \cite{Turchette00}, 
while for comparable trap dimensions and applied RF voltages, the trap depth in three-dimensional traps is considerably deeper than in two-dimensional traps. Linear blade traps, with heating rates as low as a few quanta per second and trap depths on the order of tens of eV, are known to have long ion lifetimes \cite{Rohde01, Benhelm08}.

When considering the implementation of dielectric mirrors into an ion trap, one should keep in mind the effects of dielectrics on the ion. Charges on dielectrics in vacuum are quasi-permanent and distort the ion-trap potential. They can be produced by UV light via photoelectron ionization \cite{Harlander10} in a way that is not well understood and difficult to model. The best strategy is to avoid any charging of dielectrics and to minimize the influence of possible charges on the ion. As a solution, dielectric mirrors are either placed far away from the trap \cite{Mundt02, Herskind09, Russo09, Leibrandt09, Guthoehrlein01} or the dielectric components are well shielded \cite{Steiner13, Brady11, Wilson11, VanDevender10, Kim11a}.

Therefore, when integrating an FFPC into an ion trap, the following restrictions should be respected: the ion-trap potential should be as deep as possible, and the trap geometry should be such that it shields the ion from any charges on the fibers. Furthermore, exposure of the fibers to UV light should be minimized in order to keep them from accumulating charges. In case the fibers become charged, the trap design should be flexible enough to compensate for those charges, i.e., through the application of compensation voltages. Here, we describe a setup that combines these features.

\subsection{Ion trap and vacuum chamber}

\subsubsection{Miniaturized linear Paul trap}
\label{Sec:Miniaturized_linear_Paul_trap}

\begin{figure*}
  \begin{center}
   \includegraphics[width=17cm]{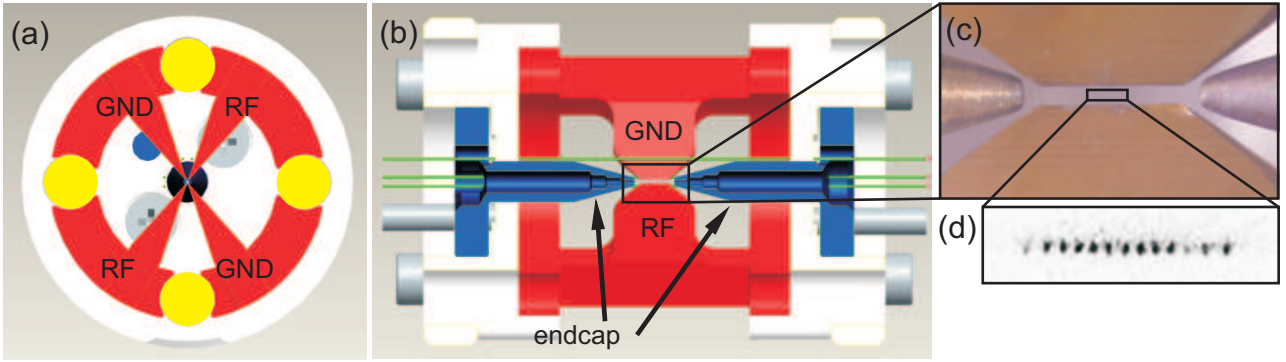}
    \caption{\label{fig:ion_trap}(a,b) Miniaturized ion-trap design. Red: two radio-frequency (RF) and two ground (GND) blades of the linear Paul trap; the distance between two opposing blades on the diagonal is $340~\mu$m (the distances between neighboring blades are $290~\mu$m and $150~\mu$m). Blue: endcap electrodes, $2.8$~mm apart. Green: four compensation electrodes, $1.7$~mm from the trap center. White: Ceramic (MACOR) mount. Yellow: Holes in which alignment rods are temporarily inserted. The plane for optical access, including laser cooling, manipulation, and fluorescence detection, is perpendicular to the fiber-cavity axis. $300~\mu$m wide holes in the endcaps provide additional optical access in this plane. (c) Photo of the ion-trap center. (d) CCD camera image of a linear string of ions.}
  \end{center}
\end{figure*}

We choose a miniaturized linear Paul trap similar to the standard design described in Refs.~\cite{Gulde03a, Riebe05a}; see Fig.~\ref{fig:ion_trap}. Four blade-shaped electrodes operated with radio frequency (RF) and ground (GND) voltages confine the ion radially, and two tip electrodes with positive voltage add confinement along the trap axis. The trap has a deep trapping potential of several electron volts inherent to three-dimensional traps, but in contrast to traps of similar design \cite{Gulde03a, Riebe05a, Schulz08}, it is miniaturized in order to make its dimensions comparable to those of the FFPCs, thus shielding the ions from charges on the fibers. The distance between opposing blade tips on the diagonal is $340~\mu$m, which means that the minimum ion-electrode distance is only $170~\mu$m; in contrast, in the design of Ref.~\cite{Riebe05a} the ion-electrode distance is $800~\mu$m. The distance along the trap axis between the two tip electrodes is $2.8$~mm, about half the length of previous designs. These axial electrodes have $300~\mu$m diameter holes for optical access along the trap axis. Another significant change is that the angle between neighboring blade electrodes is not $90^{\circ}$. In order to provide space for the fibers, the two angles between the blades shielding the fibers are increased to $120^{\circ}$. As a result, the other two angles are $60^{\circ}$. This change does not alter the trap depth significantly. The trap has four additional rod-like electrodes, parallel to the trap axis, $1.7$~mm from the trap center, and of $200~\mu$m diameter. These electrodes allow for compensation of ions' micromotion. The rods are supplied with independent voltages, enabling compensation for charges on the dielectric fibers. In Sec.~\ref{sec:simulations}, we show a simulation of the ion-trap potential.

Precise machining and positioning of the trap electrodes is necessary for a trap of such small electrode separations. The stainless-steel blade electrodes are aligned and mounted via two precision-machined glass-ceramic (MACOR) holders. The fabrication tolerance for the ceramic mounts is less than $50~\mu$m. The dimensions of the holders are then measured after machining, and the blade electrodes are subsequently electron-discharge machined to fit the holders exactly. Precision alignment of the blades with respect to the ceramic mount and to one another is done with alignment rods, which are later removed. After mounting the electrodes, we measured the dimensions of the ion trap using a microscope. We find that the inaccuracy in blade-to-blade separation is less than $30~\mu$m.

As the ion-electrode distance is very small, we expect higher ion-heating rates in comparison with larger traps, so that it may be difficult to work with ions in the motional ground state. However, the advantages of this design are manifold: the trap has a deep trapping potential, while the electrode distances are comparable to those of microfabricated traps and to the size of the FFPC. Small traps do not need RF voltages as high as those of large traps and are driven by simple RF resonators \cite{Gandolfi12}. The blade separation along the FFPC axis is only $150~\mu$m, thus shielding the ions. The small diameter of the holes in the tip electrodes helps us to align laser beams on the ion. We have successfully loaded strings of ions and single ions in the ion trap (Fig.~\ref{fig:ion_trap}(d)).

\subsubsection{Vacuum vessel}

\begin{figure}
  \begin{center}
   \includegraphics[width=8.5cm]{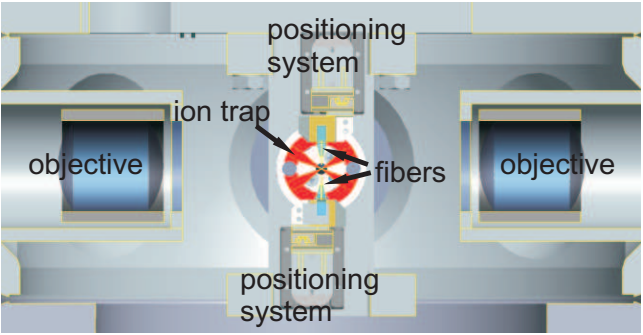}
    \caption{\label{fig:vacuum_chamber} Technical drawing of the vacuum vessel and the ion-trap fiber-cavity apparatus.}
  \end{center}
\end{figure}

To minimize collisions of ions with background gas, the trap needs to be mounted under ultra-high vacuum. The chamber is designed to optimize vacuum conditions, optical access, and stability requirements. The implementation of the FFPC leaves only one plane of optical access available for lasers and collection of fluorescence from the ions. Therefore, we chose an octagonal vacuum chamber, which provides optical access from eight sides in this plane. Fig.~\ref{fig:vacuum_chamber} shows a technical drawing of the experimental chamber. The axis orthogonal to both the FFPC and the trap axes is used for fluorescence detection. Here, within two inverted viewports, high NA objectives are installed which allow for efficient light collection and thus fast detection of the state of the ion with a camera and a PMT.

The FFPC fibers are fed into vacuum with a home-made fiber feedthrough comprising stainless-steel tubes brazed into a CF-flange. The inner diameter of the tubes is $0.5$~mm, and the fibers are glued into the tubes with vacuum epoxy\cite{EPOTEK}. In contrast to commercial fiber feedthroughs, a home-made feedthrough is advantageous as it is compatible with any fiber type, including non-standard cladding diameters.

The trap is mounted together with the FFPC on the top flange of the vacuum chamber, which also supports all electrical feedthroughs, the fiber feedthrough, and the calcium oven. A vibration-isolating material\cite{DuPont} is sandwiched between that top flange and the trap mount. The vacuum chamber sits within a hole in an optical breadboard, which allows us to use short mounting posts for optical components, thus providing improved stability.

\subsection{Integration of fibers}

The FFPC needs to be aligned with respect to the ion trap such that the center of the ion trap overlaps with the waist of the cavity mode. Furthermore, the fiber mirrors need to be aligned precisely to form a Fabry-Perot resonator, and the cavity length has to be actively stabilized.

In the FFPC experiment of Ref.~\cite{Colombe07}, it was possible to shift the trapping potential for neutral atoms with respect to the cavity using currents on the chip. The alignment of the fiber mirrors was done before the system was placed under ultra-high vacuum: one fiber was glued on a cavity mount carefully positioned with respect to the atom chip, and the second fiber was then aligned to form an FFPC with the first fiber and glued with UHV epoxy. While the glue cured, the fiber was continuously aligned by optimizing the cavity transmission signal.

We have tested the method described above to implement an FFPC into the ion trap. We find that with short cavities of about $70~\mu$m in length, this method is successful. Unfortunately, long cavities need to be aligned with considerably higher precision, and small drifts of the fiber mirrors lead to misalignment of the cavity. Although it is possible to cure the glue while keeping the FFPC aligned, the cavity signal degrades over time after the glue has set. Furthermore, the cavity signal disappears when the cavity assembly is moved or rotated. These effects may be caused by slight temperature changes or by tension in the cavity mount, and we expect that vacuum baking would also contribute to misalignment. It may be possible to stabilize the fiber mirror position passively well enough to maintain alignment, but instead we decided on an active technique to mount the FFPC inside vacuum.

\begin{figure}
  \begin{center}
   \includegraphics[width=8.5cm]{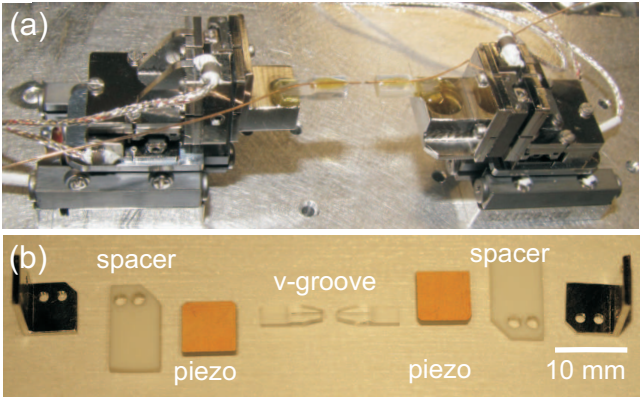}
    \caption{\label{fig:fibercavity_mount}FFPC positioning system. (a) 3D nanopositioning setup for each fiber. (b) Each fiber is glued to a glass v-groove and mounted on a shear-mode piezoelectric crystal which allows active length stabilization of the cavity. The piezoelectric crystals are glued to insulating MACOR spacers which are screwed onto the nanopositioning stages.}
  \end{center}
\end{figure}

Each fiber is mounted on a three-axis nanopositioning system compatible with ultra-high vacuum. Along the cavity axis, the fibers can be translated by up to $8$~mm\cite{SMARACT}. Along the other two axes, smaller positioners provide $4$~mm of traveling range\cite{SMARACTb}. The positioners operate via the slip-stick principle and have a minimum step size of $50$~nm, but can be moved with sub-nanometer resolution by charging the piezo actuators. The three-axis system for each fiber has dimensions of $(17\times 22\times 21)$~mm. Fig.~\ref{fig:fibercavity_mount}(a) shows a test setup of an FFPC aligned with the positioning system.

\begin{figure}
  \begin{center}
   \includegraphics[width=8.5cm]{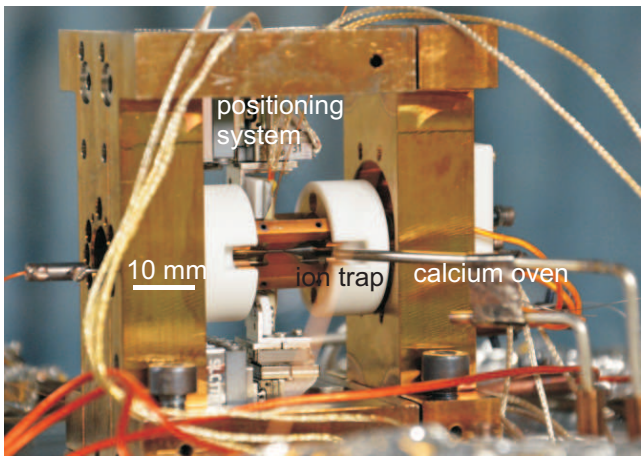}
    \caption{\label{fig:setup_foto}Photo of the linear Paul trap with an integrated FFPC. The trap axis is horizontal, and the fiber-positioning systems are visible above and below the trap. The calcium oven points towards the trap center from the right side. Vacuum-compatible coaxial cables connect positioners and trap electrodes to vacuum feedthroughs.}
  \end{center}
\end{figure}

Each fiber is glued to a Pyrex v-groove, which sits on a shear-mode piezoelectric actuator; see Fig.\ref{fig:fibercavity_mount}(b). The additional actuator is needed to stabilize the cavity length actively as the bandwidth of the positioning system is too low. These actuators are fixed to the positioners, and the whole assembly is mounted on the same holder as the ion trap.
Fig.~\ref{fig:setup_foto} shows a photo of the ion-trap setup together with the FFPC aligned via the micropositioning system.

The in-vacuum positioning system offers a range of advantages for the setup. First, it provides the option of realigning the cavity under vacuum in case of misalignment due to baking or transport of the vacuum chamber. Second, it is possible to pull the fibers back by almost a centimeter from the trapping region, allowing the trapping of ions without dielectrics close to the trapping region. Also, if the fibers are out of the trapping region during ion loading, the trap blades shield the fibers, and the fiber mirrors do not get coated with calcium. Furthermore, the fibers can be moved towards the trap center iteratively while compensating for charges on the dielectric mirrors via the compensation electrodes. Finally, the positioners allow us to change the mirror separation of the FFPC inside vacuum and thus build cavities of variable cavity length and waist, resulting in an adjustable coupling parameter $g$.

\subsection{Simulations}
\label{sec:simulations}

The trap potential of the Paul trap is characterized in numerical simulations using CPO\cite{CPO} and Matlab. CPO solves the electromagnetic field equations for each electrode, which are then combined with Matlab to give the net potential over the trapping region. The fibers are included in the simulations as dielectric cylinders. These simulations were used to determine trap depth and frequencies and to optimize the trap geometry.

\begin{figure*}
  \begin{center}
    \includegraphics[width=17cm]{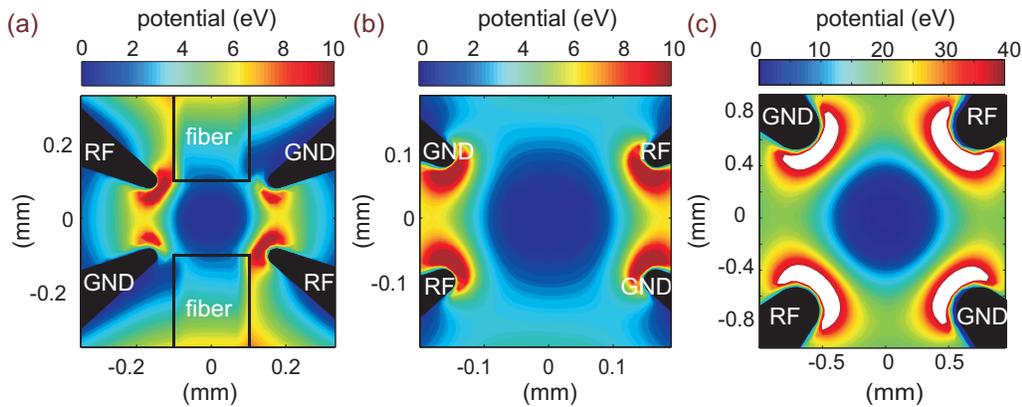}
    \caption{\label{fig:simulated_trap_potential} (a) Simulation of the trap potential of the miniaturized linear Paul trap. The radio-frequency (RF) and ground (GND) blades as well as the dielectric fibers are indicated in black. The potential in the plane perpendicular to the trap axis is plotted in eV. The fibers are separated by $200~\mu$m. (b) Trap potential without fibers. (c) Trap potential of a symmetric trap with $90^{\circ}$ angles between RF and ground blades and blade separations of $1.6$~mm.}
  \end{center}
\end{figure*}

We calculate the trap potential of the trap described in Sec.~\ref{Sec:Miniaturized_linear_Paul_trap} for the following parameters: RF amplitude of $130$~V, RF frequency of $2\pi\times35$~MHz, and tip electrodes at $200$~V. As we cannot predict the amount of charge on the fibers, we assume it to be zero in the simulations. Fig.~\ref{fig:simulated_trap_potential}(a) shows the potential in the radial direction. The radial trap frequencies are $2\pi\times9.9$~MHz and $2\pi\times9.6$~MHz, the axial trap frequency is $1.9$~MHz, and the trap depth is $1.3$~eV.

In Fig.~\ref{fig:simulated_trap_potential}(b), the potential of the same trap  without fibers is shown. In the absence of the dielectric fibers,  the trap depth is now $2$~eV instead of $1.3$~eV.

Changing the blade angle from the asymmetric case of $120^{\circ}$ and $60^{\circ}$ between RF and ground blades to angles of $90^{\circ}$ results in a increase of trap depth of the radio frequency potential by a factor of three. At the same time, however, the influence of the DC endcap potential decreases, and therefore the overall trap depth does not change significantly. The radial symmetry of the trap is no longer broken and the radial trap frequencies are degenerate.

The trap implemented in the experimental setup has blade separations that are five times smaller than those in the trap of Ref.~\cite{Riebe05a}. To maintain a stable trapping configuration in the smaller trap, either the radio-frequency amplitude has to be decreased or its frequency has to be increased. Both result in a lower trap depth: the depth of the miniaturized trap is an order of magnitude smaller. Fig.~\ref{fig:simulated_trap_potential}(c) shows the potential of the trap from Ref.~\cite{Riebe05a}, in which the blade separations along the diagonal are $1.6$~mm, all angles between RF and ground blades are $90^{\circ}$, and the trap depth is $20$~eV.

The simulations confirm that the trapping potential of the trap implemented in the setup has suitable trap depth and trap frequencies. Although not as deep as a conventional Paul trap, it is nevertheless an order of magnitude deeper than typical planar ion traps, giving this geometry a clear advantage over planar designs.

\section{Summary and outlook}

We have built an ion-trap apparatus with an integrated FFPC of tunable length and thus of tunable coupling parameter $g$. In the process, we have set up and tested a miniature linear Paul trap for $^{40}$Ca$^+$. Furthermore, we have extended the parameter regime for CO$_2$-laser shaped fiber tips in order to build high-finesse FFPCs of up to $350~\mu$m in length, compatible with the integration of an ion trap. We have performed several experiments to characterize the properties of FFPCs, which we expect will further the development of this new technology. We describe several techniques for handling mirror-coated fibers and discuss methods to improve the fiber-mirror finesse.

Our apparatus is intended for performing cavity QED experiments. In particular, we would like to demonstrate strong coupling between an ion and a photon, in which $g \gg \kappa,\gamma$ \cite{Kimble08a}. We have calculated that our system enters this regime for a cavity in near-concentric configuration. Similar to the cavity described in Ref.~\cite{Stute12a}, the fiber cavity can be tuned into resonance with the $P$ to $D$ transition in $^{40}$Ca$^+$, where $D$ is a metastable state used as a qubit state. With a drive laser coupling the electronic ground state $S$ and the intermediate state $P$, a vacuum-stimulated Raman transition from $S$ to $D$ produces single photons inside the cavity. An ion coupled to an FFPC would enable a coherent interaction rate $g$ that dominates over the ion's spontaneous decay rate $\gamma$, providing access to a new experimental regime for ion-trap cavity QED experiments.

\begin{acknowledgments}
The authors would like to thank Ramin Lalezari from ATFilms for valuable discussions on mirror coatings; Frederik Klauser from the Institute of Physical Chemistry of the University of Innsbruck for his aid with the XPS measurements; Stefan Haslwantler, Johannes Ghetta, Ben Ames, and Jasleen Lugani for valuable discussions and aid in the construction of the FFPCs and the experimental apparatus; and the mechanical workshop of the Institute of Experimental Physics at the University of Innsbruck.
We gratefully acknowledge support from the Austrian Science Fund (FWF):  Project. Nos. F4003 and F4019,
the European Research Council through the CRYTERION Project,
the European Commission via the Atomic QUantum TEchnologies (AQUTE) Integrating Project,
and the Institut f\"ur Quanteninformation GmbH.\\
\end{acknowledgments}

%
\end{document}